\documentclass[11pt,letterpaper]{article}

\usepackage[margin=0.7in]{geometry}
\usepackage{amsmath,amssymb}
\usepackage{graphicx}
\usepackage{subcaption}
\usepackage{fourier}
\usepackage{cite}

\usepackage{xcolor}
\definecolor{mattepink}{HTML}{D47B95}
\usepackage[colorlinks=true,allcolors=blue]{hyperref}
\usepackage{authblk}
\usepackage{titlesec}
\usepackage{fontawesome5}
\titlespacing*{\section}{0pt}{6pt}{2pt}
\begin{document}

\title{
Atmospheric Turbulence-Aware Optimization of Generalized Kennedy Receivers for Coherent-State BPSK Quantum Communication

}

\author[1]{Md Ashraf Hossain Ifty}
\author[1]{Rahul Bhadani}

\affil[1]{
\small
Electrical and Computer Engineering,
The University of Alabama in Huntsville,
Huntsville, AL 35899, USA
}

\affil[ ]{
\small
\textcolor{mattepink}{\texttt{mi1497@uah.edu}},
\textcolor{mattepink}{\texttt{rahul.bhadani@uah.edu}}
}

\date{}


\maketitle

\vspace{-1.5em}

\begin{abstract}
\noindent
We investigate generalized Kennedy receivers for coherent-state BPSK communication over turbulent free-space optical channels. An ergodic mutual-information framework is developed to jointly optimize the receiver displacement and input prior probabilities under Gamma--Gamma fading.
\end{abstract}

\vspace{0.8em}


\begingroup
\renewcommand{\thefootnote}{}
\footnotetext{This manuscript corresponds to the paper accepted for presentation at the
2026 Frontiers in Optics + Laser Science (FiO LS), Rochester, New York, USA.
A substantially expanded journal version is under preparation.}
\addtocounter{footnote}{-1}
\endgroup

\section{Introduction}

Coherent state binary phase-shift keying (BPSK) is a fundamental modulation format for photon-starved optical communication systems, including deep-space and free-space optical (FSO) links \cite{semenov2025quantum}. Structured quantum receivers such as the Kennedy and generalized Kennedy (GK) receivers offer practical architectures capable of approaching quantum-limited detection performance while remaining experimentally realizable \cite{kennedy1973,takeoka2008}. Recent studies have demonstrated that the achievable information rate of the GK receiver can be improved through joint optimization of the receiver displacement and input symbol prior distribution \cite{bhadani2020}. However, these analyses largely assume deterministic bosonic loss channels and do not account for atmospheric turbulence encountered in practical FSO systems. In realistic propagation environments, atmospheric turbulence causes random fluctuations in channel transmissivity \cite{ghalaii2022quantum}. As a result, the distinguishability of non-orthogonal coherent states degrades, forcing a shift in the optimal receiver operating point \cite{kodela2026, yuan2018free}. In this work, we develop an ergodic mutual-information framework for coherent state BPSK communication over Gamma-Gamma fading channels and jointly optimize the GK receiver displacement and symbol prior to maximize achievable information. We compare the resulting receiver-constrained information rates and optimal operating parameters against standard deterministic channel baselines. The results demonstrate that atmospheric turbulence significantly alters both the achievable information rate and the capacity-maximizing receiver operating point.

\begin{figure*}[t]
    \centering

    \begin{subfigure}[b]{0.32\textwidth}
        \centering
        \includegraphics[width=\linewidth]{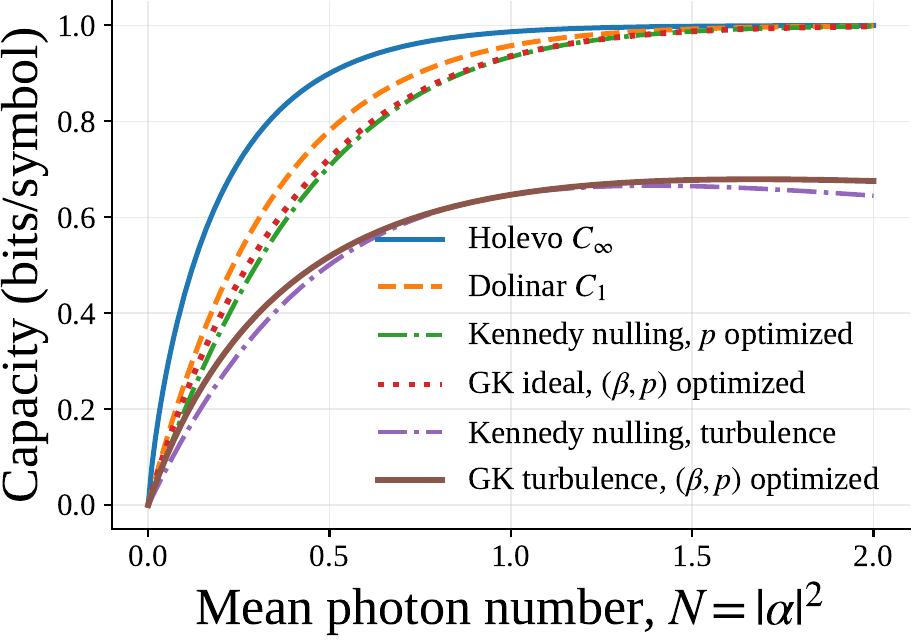}
        \caption{}
        \label{fig:capacity}
    \end{subfigure}
    \hfill
    \begin{subfigure}[b]{0.32\textwidth}
        \centering
        \includegraphics[width=\linewidth]{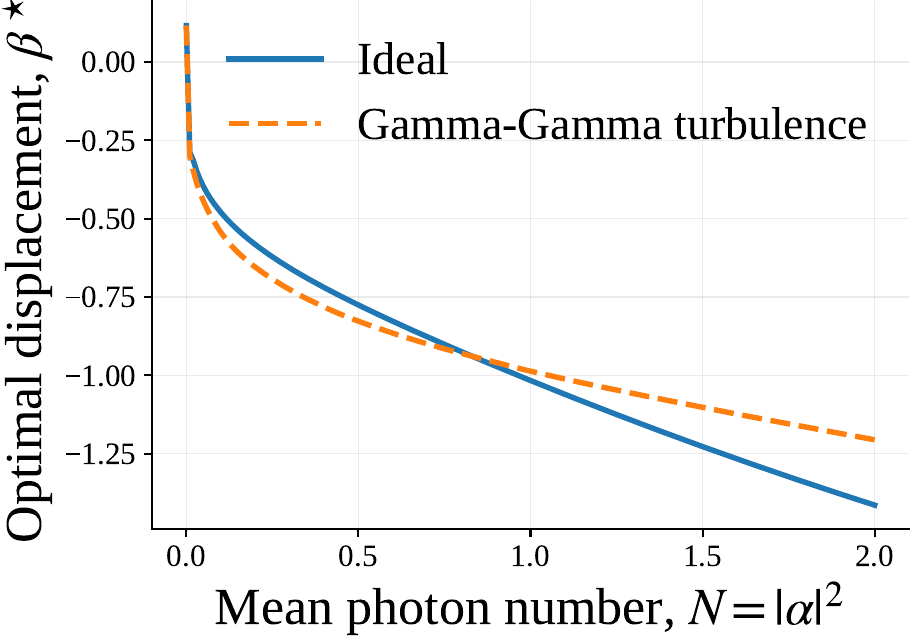}
        \caption{}
        \label{fig:optimal_beta}
    \end{subfigure}
    \hfill
    \begin{subfigure}[b]{0.32\textwidth}
        \centering
        \includegraphics[width=\linewidth]{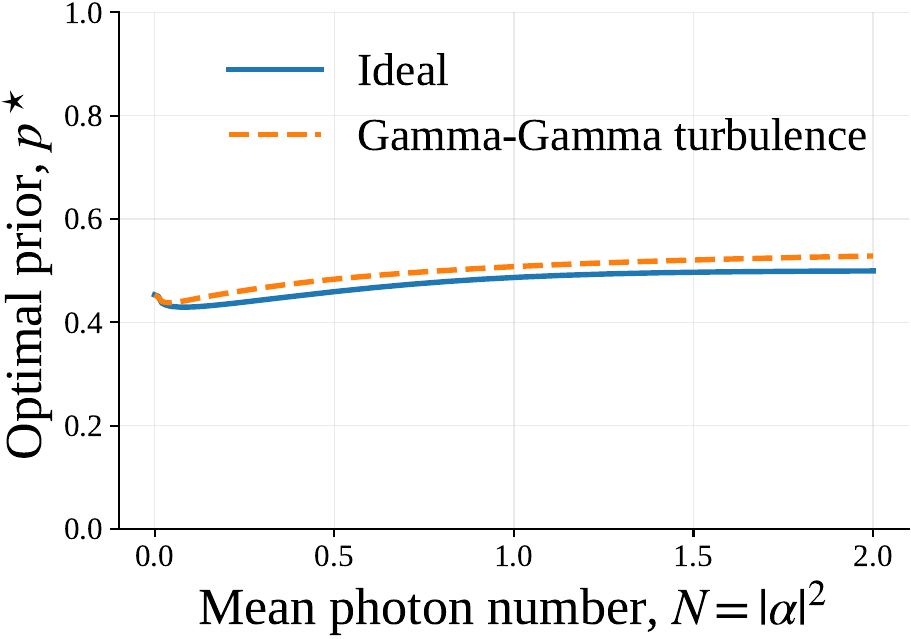}
        \caption{}
        \label{fig:optimal_prior}
    \end{subfigure}

    \caption{
    Performance of the generalized Kennedy receiver under deterministic and
    Gamma--Gamma fading channels as a function of mean photon number.
    (a) Achievable information rate compared with the Kennedy nulling receiver,
    Dolinar (Helstrom) limit, and Holevo limit.
    (b) Capacity-maximizing displacement parameter.
    (c) Capacity-maximizing source prior probability.
    }
    \label{fig:results}
\end{figure*}

\section{System Model and Ergodic Information Analysis}

We consider coherent state BPSK signaling with alphabet $\mathcal{X}=\{|-\alpha\rangle,|+\alpha\rangle\}$, mean photon number $N=|\alpha|^2$, and input prior probabilities $P(X=0)=p$ and $P(X=1)=1-p$. Atmospheric turbulence is modeled as a Gamma-Gamma fading process with random channel transmissivity denoted by a random variable $T$. Let $\tau$ denote a realization of $T$. Conditioned on $\tau$, the pure-loss channel transforms the transmitted coherent states according to $|\pm\alpha\rangle \rightarrow |\pm\sqrt{\tau}\alpha\rangle$ \cite{al2001mathematical, andrews2005laser}. The turbulence process is assumed to be quasi-static, such that the channel transmissivity remains constant over a symbol interval and varies independently between realizations. Furthermore, the receiver is assumed to possess channel state information during mutual-information evaluation, allowing the received coherent states to remain conditionally pure. The Gamma-Gamma channel is modeled as the product of two independent Gamma random variables, $T=UV$, where $U\sim\Gamma(\alpha_g,1/\alpha_g)$ and $V\sim\Gamma(\beta_g,1/\beta_g)$ \cite{geng2025gpu}. The scale parameters are selected such that $\mathbb{E}[U]=\mathbb{E}[V]=1$, yielding $\mathbb{E}[T]=1$ and allowing turbulence-induced fluctuations to be analyzed independently of deterministic path loss. At the receiver, a GK receiver applies a real-valued coherent displacement $\beta$ followed by ideal on-off photon detection. The displaced states become $|\beta-\sqrt{\tau}\alpha\rangle$ and $|\beta+\sqrt{\tau}\alpha\rangle$. Since the probability of a no-click event for a coherent state $|\gamma\rangle$ is $P(Y=0|\gamma)=|\langle0|\gamma\rangle|^2=e^{-|\gamma|^2}$, the conditional no-click probabilities are given by $q_0(\tau)=e^{-(\beta-\sqrt{\tau}\alpha)^2}$ and $q_1(\tau)=e^{-(\beta+\sqrt{\tau}\alpha)^2}$. The corresponding marginal no-click probability is $\bar q(\tau)=pq_0(\tau)+(1-p)q_1(\tau)$. The resulting receiver induces a binary-input binary-output asymmetric channel whose conditional mutual information is given by 
$
I(X;Y|\tau)=h_2(\bar q(\tau))-p\,h_2(q_0(\tau))-(1-p)\,h_2(q_1(\tau)),
$ where $h_2(x)$ denotes the binary entropy function. For the deterministic channel, corresponding to $\tau=1$, the achievable information rate of the fully optimized GK receiver is obtained from $C_{\mathrm{GK}}(N)=\max_{\beta,p} I(X;Y)$ \cite{goldsmith1997capacity}. Under Gamma-Gamma turbulence, the achievable rate is characterized by the ergodic mutual information $C_{\mathrm{GK}}^{\mathrm{GG}}(N)=\max_{\beta,p}\mathbb{E}_T[I(X;Y|T)]$. Since the expectation with respect to the Gamma-Gamma fading distribution does not admit a closed-form analytical expression, it is evaluated numerically using Monte Carlo integration according to $\mathbb{E}_T[I(X;Y|T)] \approx \frac{1}{K}\sum_{k=1}^{K}I(X;Y|T_k)$, where $T_1,\ldots,T_K$ are independent realizations drawn from the Gamma-Gamma distribution with parameters $(\alpha_g,\beta_g)$ \cite{metropolis1949monte}. By the Law of Large Numbers, this estimator converges to the true expectation as $K\rightarrow\infty$. To quantify the impact of atmospheric fading, we consider both a Kennedy nulling baseline and a fully optimized GK receiver. The Kennedy nulling receiver fixes the displacement to $\beta_{\mathrm{null}}=\sqrt{N}$ and optimizes only the source prior, whereas the GK receiver jointly optimizes $(\beta,p)$ to maximize the achievable information rate. Performance is benchmarked against the Dolinar (Helstrom) limit, $C_1(N)=1-h_2(P_e)$ with $P_e=\frac{1}{2}(1-\sqrt{1-e^{-4N}})$, and the Holevo limit, $C_\infty(N)=h_2\!\left(\frac{1-e^{-2N}}{2}\right)$, which represent the ultimate single-symbol and collective-measurement quantum information bounds, respectively.

\section{Results and Discussion}

The proposed framework was evaluated for coherent state BPSK signaling over both deterministic and Gamma-Gamma fading channels with turbulence parameters $(\alpha_g,\beta_g)=(4,2)$. For each mean photon number, the displacement parameter and source prior were jointly optimized to maximize the achievable information rate.

Figure~\ref{fig:results} summarizes the impact of atmospheric turbulence on the achievable information rate and optimal receiver parameters. As shown in Figure~\ref{fig:capacity} the jointly optimized GK receiver consistently outperforms the Kennedy nulling baseline under both deterministic and Gamma-Gamma fading conditions. While the optimized receiver approaches the Dolinar benchmark at higher mean photon numbers, it remains below the Holevo limit due to its receiver-constrained architecture. Atmospheric turbulence reduces the distinguishability of the received coherent states, introducing a noticeable capacity penalty. Despite this degradation, jointly optimizing the displacement and source prior consistently yields a measurable performance gain across all operating ranges. Figure~\ref{fig:optimal_beta} shows the optimal displacement parameter as a function of mean photon number. The capacity-maximizing displacement is negative across most of the operating range and becomes increasingly negative as the mean photon number increases. This behavior indicates that the mutual-information optimal operating point differs significantly from the conventional Kennedy nulling condition. Under Gamma-Gamma fading, the optimal displacement remains closer to zero than in the deterministic channel. This trend suggests that turbulence favors a more conservative receiver configuration to maintain robustness against random channel fluctuations. Figure~\ref{fig:optimal_prior} presents the optimal source prior probability. In the low mean photon number regime, the optimal prior deviates slightly from the symmetric value of $p^\star=0.5$. However, it gradually approaches an equal-probability signaling strategy as the coherent states become more distinguishable. As state distinguishability improves, the signaling strategy steadily converges toward equal probabilities. Nevertheless, the turbulence-aware prior stays consistently higher than the baseline ideal case, demonstrating that atmospheric fading induces further structural asymmetry within the channel. However, the relatively small separation between the prior curves compared with the displacement curves suggests that receiver adaptation under turbulence is driven primarily by displacement optimization rather than by modifications to the input distribution. Future work will extend the proposed framework to higher-order modulation formats, including $M$-ary phase-shift keying and quadrature amplitude modulation, to evaluate the impact of atmospheric turbulence on larger signal constellations. Additionally, non-stationary turbulence models will be investigated to characterize how time-varying channel statistics influence receiver optimization and achievable information rates. 

\vspace{1em}
\noindent\textbf{Code Availability:} The complete simulation source code is openly available on \ \href{https://github.com/ashraf711/gk-fso-turbulence-capacity/tree/main}{\texttt{GitHub}} \faGithub.

\bibliographystyle{IEEEtran}
\bibliography{ref}
\end{document}